# Temperature-dependent Fowler-Nordheim electron barrier height in $SiO_2$/4H-SiC MOS capacitors

Patrick Fiorenza[1*], Marilena Vivona[2], Ferdinando Iucolano[3], Andrea Severino[3], Simona Lorenti[3], Giuseppe Nicotra[1], Corrado Bongiorno[1], Filippo Giannazzo[1], Fabrizio Roccaforte[1]

[1]*Consiglio Nazionale delle Ricerche – Istituto per la Microelettronica e Microsistemi (CNR-IMM), Strada VIII, n.5 Zona Industriale, I-95121 Catania, Italy*
[2] *Optoelectronics Research Centre University of Southampton, United Kingdom*
[3] *STMicroelectronics, Stradale Primosole 50, I-95121 Catania, Italy*
**patrick.fiorenza@imm.cnr.it*

## Abstract

This paper reports on the physical and temperature-dependent electrical characterizations of the oxide/semiconductor interface in MOS capacitors with a $SiO_2$ layer deposited on 4H-SiC using dichlorosilane and nitrogen-based vapor precursors. The capacitors, subjected to a standard post deposition annealing process in $N_2O$, exhibited an interface state density $D_{it} \approx 9.0\times10^{11}cm^{-2}eV^{-1}$ below the conduction band edge. At room temperature, a barrier height (conduction band offset) of 2.8 eV was observed, along with the presence of negative charges in the insulator. The $SiO_2$ insulating properties were evaluated by studying the experimental temperature-dependence of the gate current. In particular, the temperature-dependent electrical measurements showed a negative temperature coefficient of the Fowler-Nordheim electron barrier height ($d\Phi_B/dT = -0.98$ meV/°C), which was very close to the expected value for an ideal $SiO_2$/4H-SiC system and much lower compared to the values reported for thermally grown $SiO_2$. This smaller dependence of $\Phi_B$ on the temperature and the increase of the current level with temperature in the transcharacteristics measured in the relative fabricated MOSFETs represents a clear advantage of our deposited $SiO_2$ for the operation of MOSFET devices at high temperatures.

## 1. Introduction

Thanks to its exceptional physical properties, 4H-SiC is an excellent wide band gap (WBG) semiconductor for high power electronics applications [1]. In particular, power switches based on 4H-SiC metal-oxide-semiconductor field effect transistors (4H-SiC MOSFETs) can be used in energy conversion systems for automotive and renewable energies, providing a significant improvement in terms of energy efficiency. Clearly, studying the properties of the $SiO_2$/4H-SiC interfaces (interface state density $D_{it}$, charge trapping mechanisms, conduction mechanisms through the insulator, etc.) is particularly important, as this system represents the core of the MOSFET device. In fact, it is well known that a high $D_{it}$ or the presence of near-interface-traps (NITs) can be detrimental both for the channel mobility and for the threshold voltage stability of the devices [2,3]. In addition, reliability issues play a critical role for 4H-SiC MOSFETs, where the reduced band offset can induce leakage current enhancement and the consequent early degradation of the gate insulator at high electric fields. In this context, the full understanding of the mechanisms governing the carrier transport through the gate oxides and the dielectric degradation is still missing and currently under debate.[4]

In literature, $SiO_2$-gate oxide in 4H-SiC MOSFETs could be formed either by thermal oxidation [5] or by the deposition from vapor phase techniques at moderate temperatures (e.g., TEOS at 400°C [6]). In practice, post oxidation annealing (POA) or post deposition annealing (PDA) treatments are required to reduce the $D_{it}$ and to improve the MOSFETs channel mobility. In the case of thermally grown oxides, some authors reported the presence of a disordered C-rich interfacial layer [7,8,9], detected even after nitridation annealing in NO [10], which can penalize also the MOS-stack reliability. In particular, the conduction band offset between $SiO_2$/4H-SiC system, lower compared to the $SiO_2$/Si system, deserves particular attention because the enhanced leakage current can induce the early wear out of the insulating layer and the premature power device failure. Recently, it has been demonstrated the presence of active defects in thermally grown oxides onto n-type 4H-SiC that enhance the current conduction through the insulator in MOS capacitor [11]. The Poole-Frenkel (PF) emission is believed to be responsible for this current enhancement with respect to the theoretical bottom limit of the Fowler-Nordheim tunneling. Specifically, in order to avoid the formation of C-related defects, which causes an increase of the gate current in thermal oxides, the use of deposited oxides has been explored [12,13,14,15].

Hence, studying the temperature dependence of the Fowler–Nordheim current in 4H-SiC MOS systems is of considerable interest because of the practical implications in the evaluation of the high-field and high-temperature performance of both MOSFETs and Insulated-Gate-Bipolar-Transistors (IGBT's) [16].

In this work, a temperature dependent electrical characterization of 4H-SiC MOS capacitors, with the oxide layer deposited at high temperature (800°C) from dichlorosilane and nitrogen-based precursors and subjected to PDA in $N_2O$, is presented. Using the standard Fowler-Nordheim (FN) tunneling formulation the electron barrier height could be determined from room temperature up to 150°C. In particular, the temperature coefficient of the FN electron barrier height ($d\Phi_B/dT$ vs T) was considered to evaluate the quality of the deposited oxide with respect to the ideal $SiO_2$/4H-SiC system. The results showed a nearly-ideal thermal behavior of our gate oxide, as the temperature coefficient of the FN barrier (– 0.98 meV/°C) was very close to the ideal one and much lower compared to that reported for thermally grown $SiO_2$ layers.

## 2. Experimental Details

The $SiO_2$ layer was deposited by low pressure chemical vapor deposition (LPCVD) on a n-type 4H-SiC epitaxial sample with doping level of $N_D=1\times10^{16}$ cm$^{-3}$. The deposition process of $SiO_2$ was carried out at a temperature of about 800 °C, using dichlorosilane $SiH_2Cl_2$, (DCS) and $N_2O$ as gaseous precursors. After the deposition of the $SiO_2$-layer, the sample was subjected to a standard post-deposition annealing (PDA) treatment in $N_2O$ ambient at 1150°C for 4 hours [17].

Then, MOS capacitors were fabricated using a heavily doped n-type poly-Si as gate electrode and Nickel as back contact. Firstly, physical analysis of the oxide by transmission electron microscopy (TEM) and secondary ion mass spectrometry (SIMS) were performed on blanket samples. Then, electrical capacitance-voltage (C-V), conductance-voltage ($G_p/\omega$-V) and current-voltage (I-V) measurements were performed on the MOS capacitors by a Cascade Summit 12000M probe station. These measurements were carried out from room temperature up to 150°C.

## 3. Results and Discussion

First of all, cross section TEM analysis of the deposited oxide subjected to PDA has been carried out. As can be seen in Fig. 1a, the deposited $SiO_2$ is conformal to the 4H-SiC substrate showing a total thickness of about 49 nm.

Fig. 1b reports the SIMS profile *of Silicon,* Carbon, Oxygen and Nitrogen in the oxide/semiconductor interfacial region. The N-concentration, originated from the PDA treatment in $N_2O$, has been extrapolated directly from the SIMS analysis and results in the order of $10^{21}$ atoms/cm$^3$. As matter of fact, the N concentration is very focused at that interface as generally the post oxidation annealing produces a migration of N atoms from the top surface through the oxide layer with N atoms segregating at the $SiO_2$/SiC interface. Thus, it is the areal density related to the N-peak that becomes an interesting parameter. From a mathematical integration of the volumetric density over a spatial region at the interface related to the N-peak, we obtain a N-areal density of about $2\times10^{14}$ atoms/cm$^2$.

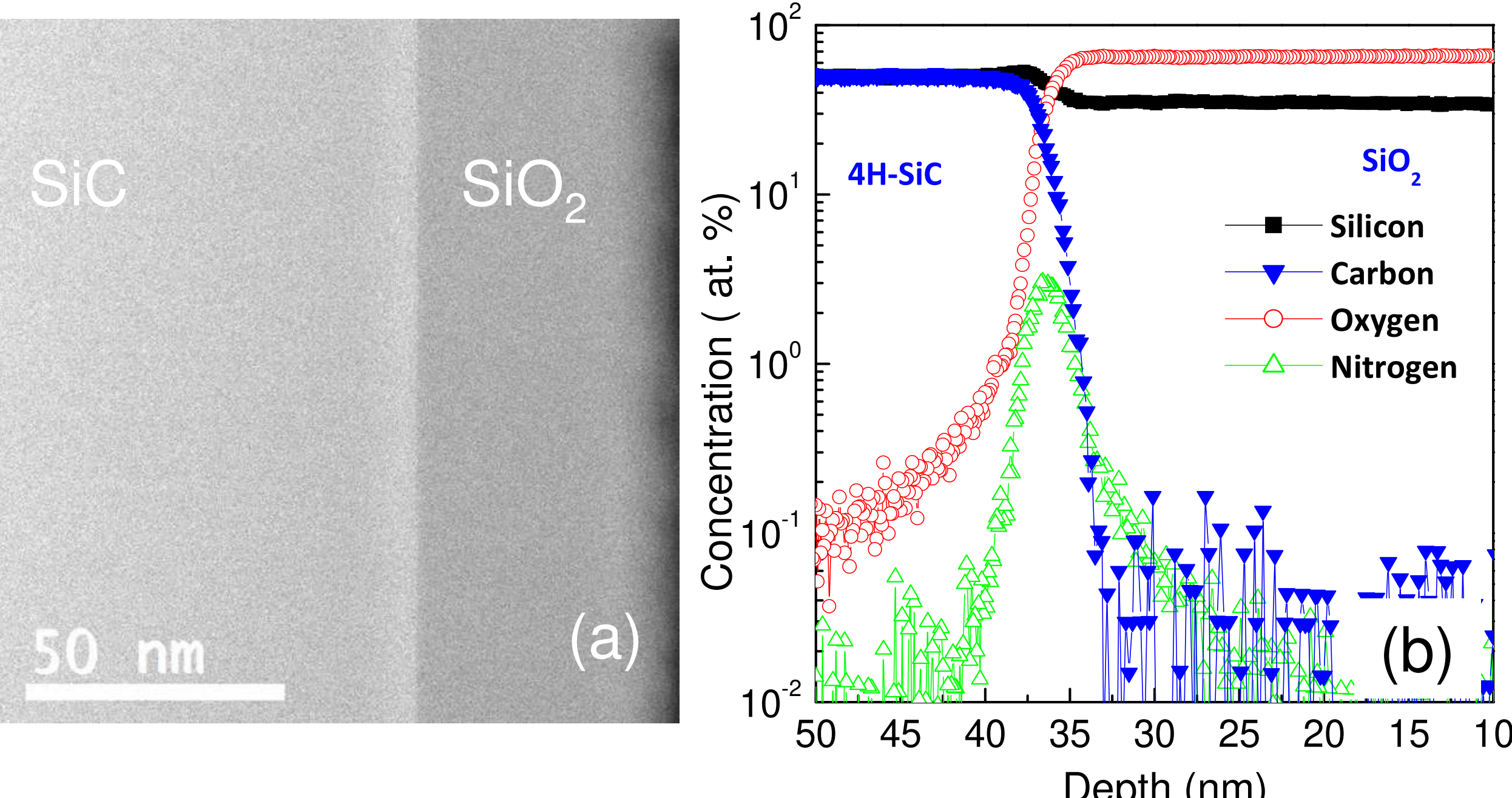

Fig. 1. (a) TEM cross section and (b) SIMS profiles acquired in the 4H-SiC/$SiO_2$ system under investigation.

It is worth noting that the C-profile decreases steeply towards the $SiO_2$. From this analysis, no significant C accumulation is detected at the interface. However, it cannot be ruled out that small amount of C (below the detection limit) is present inside the $SiO_2$ as a consequence of the re-oxidation due to PDA.

The electrical characterization has been carried out starting with a set of temperature dependent C-V measurements, acquired on capacitors having an area of 0.0225 cm$^2$. The temperature dependence has been evaluated in order to monitor any possible change of the flat band voltage ($V_{FB}$) and to determine with high accuracy the effective charge density. Fig.2a shows the C-V curves measured at different temperatures from 25°C up to 150°C. At room temperature, the experimental $V_{FB}$ is positively shifted with respect to the ideal C-V value due to the presence of an effective negative charge in the MOS system. Quantitatively, an experimental $V_{FB}$ = -0.26 V has been determined, which corresponds to an effective negative charge density of $Q_{eff} = 2.2\times10^{11}$ cm$^{-2}$ obtained from the comparison with the ideal $V_{FB}$ value ($V_{FB}$= -0.72 V). Moreover, as can be seen in Fig. 2a, the C-V curves are quite stable with the temperature and, as reported in Fig. 2b, the flat band voltage exhibits only a variation of -0.8mV/°C (Fig.2b) with increasing the temperature ($V_{FB}$ = -0.36 eV at 150°C).

The density of interface states $D_{it}$ was determined by means of the conductance method from $G_p/\omega$ vs frequency. Fig.3 displays the values of $D_{it}$ from the edge of the conduction band $E_C$. The value of $D_{it} \approx 9.0 \times 10^{11} cm^{-2} eV^{-1}$, obtained at 0.2 eV from $E_C$, is comparable to those typically measured in 4H-SiC MOS capacitors employing thermally grown oxides and subjected to the same post oxidation annealing (POA) ($D_{it}$ = 0.8-1.8×$10^{12} cm^{-2} eV^{-1}$) [18,19,20]. The presence of a low density of fixed charges as well as the $D_{it}$ value comparable to thermal growth process, indicate that the oxide properties are ultimately determined by the annealing treatments (POA or PDA) in $N_2O$.

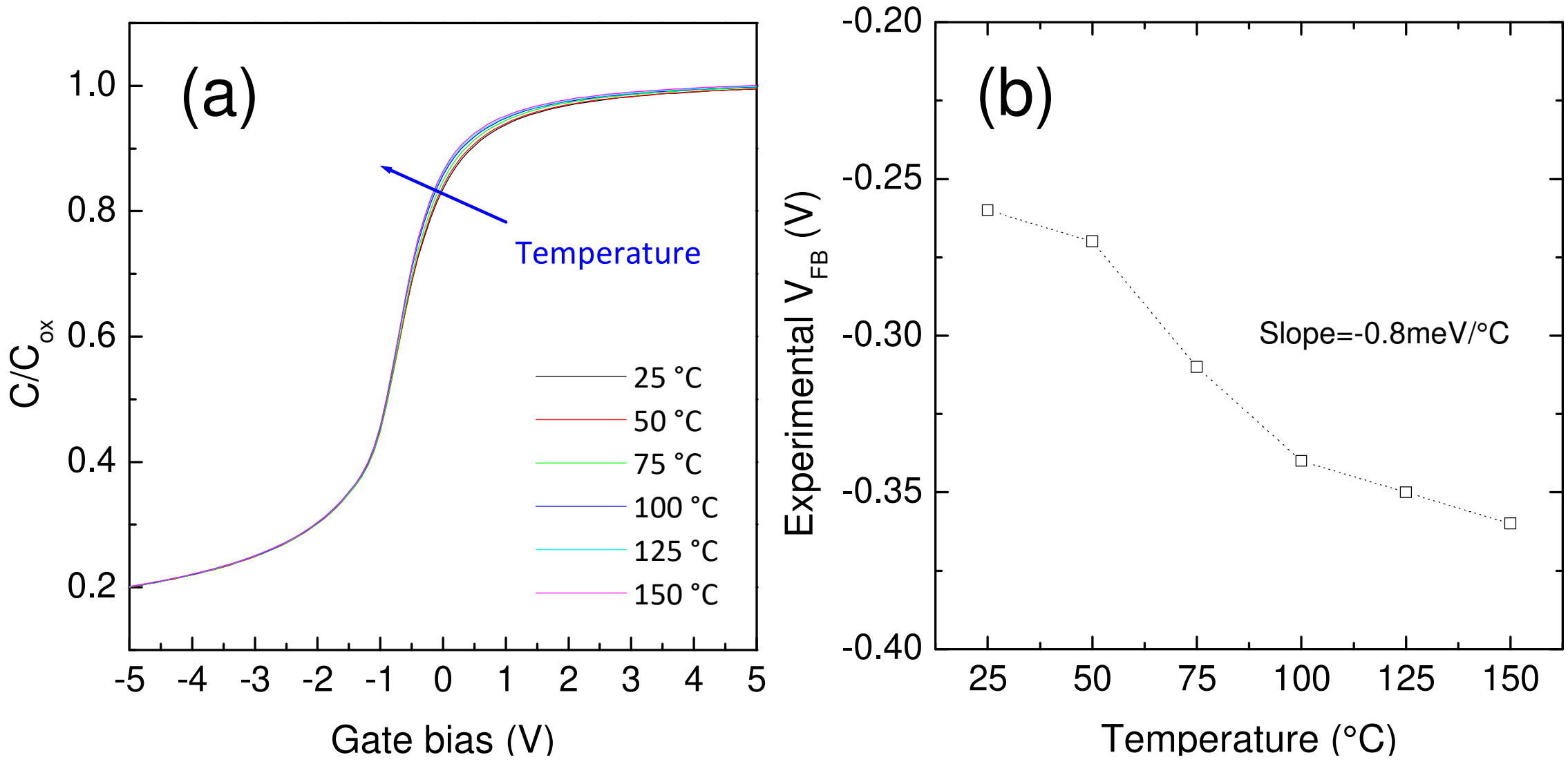

Fig. 2. (a) Temperature dependent C-V curves from room temperature up to 150°C. (b) Temperature dependence of the experimental flat band voltage $V_{FB}$.

The insulating properties were also monitored by I-V analysis, with I-V characteristics measured from room temperature up to 150°C in the MOS capacitor.

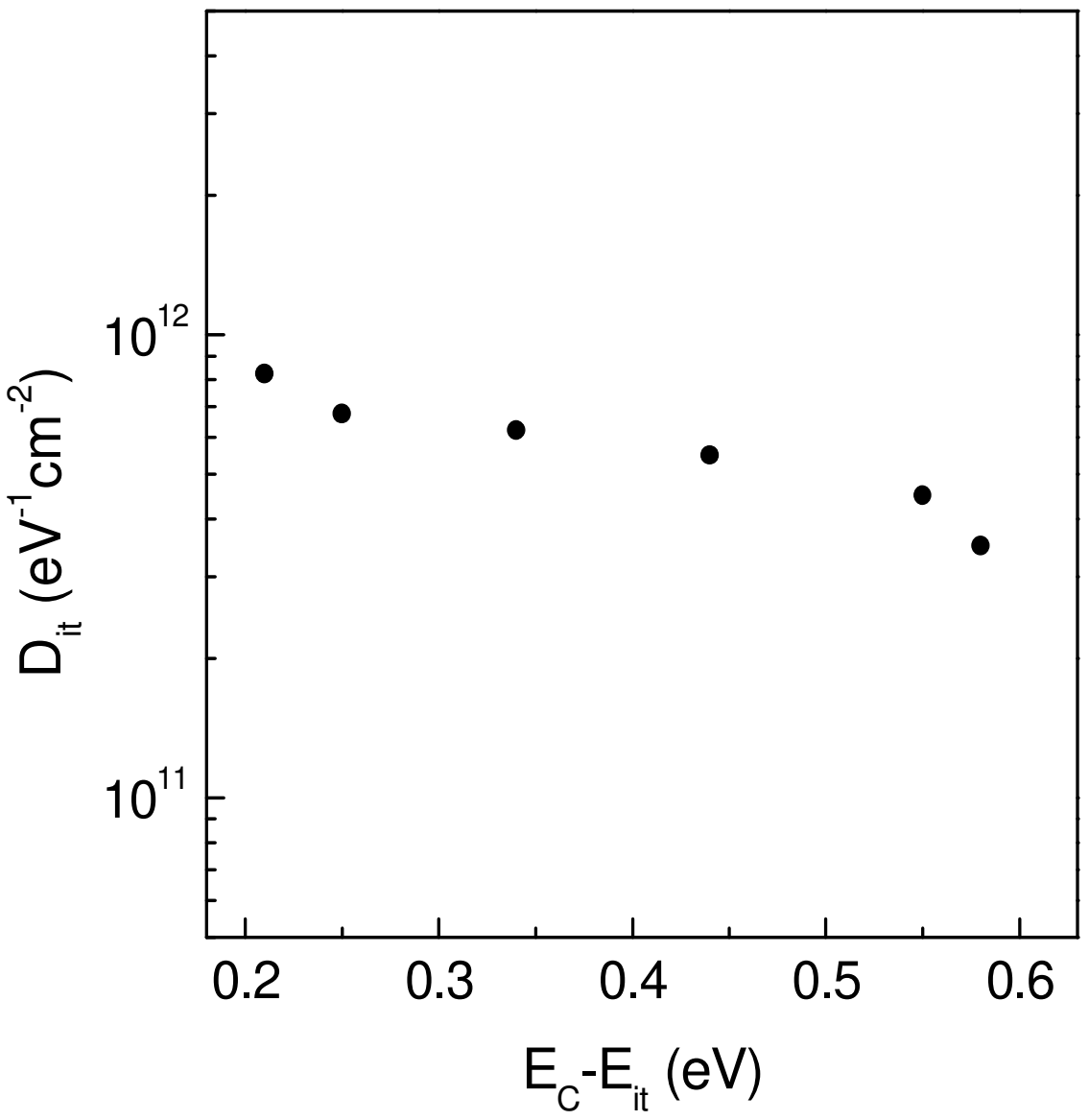

Fig. 3. Energy distribution within the gap of the interface state density $D_{it}$.

The I-V characteristics were analyzed using the Fowler-Nordheim formalism and the Lenzlinger-Snow (LS) equation:

$$ln\left(J/E_{ox}^{2}\right) = ln\left(\frac{q^{3}\left(m_{SiC}/m_{ox}\right)}{8\pi h\Phi_{B}}\right) - \frac{8\pi\sqrt{2m_{ox}\Phi_{B}^{3}}}{3qh}\frac{1}{E_{ox}} \qquad (1)$$

where J is the current density, $E_{ox}$ is the electric field across the oxide, $m_{SiC}$ and $m_{ox}$ are the effective electron masses in the SiC substrate and in the insulator respectively, q is the electron charge, h is the Plank constant and $\Phi_B$ is the tunnelling barrier height for electrons [21]. By the fits in the linear region of the *"FN plots"* $\ln(J/E_{ox}^2)$ vs $1/E_{ox}$ shown in Fig. 4, it was possible to determine the values of $\Phi_B$ for electrons as a function of the temperature (Fig. 5). It has to be emphasized that the current measurements are performed varying the gate bias ($V_G$). However, we have considered the total electric field across the insulator, taking into account also the T-dependence of the $V_{FB}$. This can be written as:

$$E_{ox}\left(V_G;T\right) = \frac{V_G - V_{FB}^{exp}\left(T\right)}{t_{ox}} \qquad (2)$$

where $t_{ox}$ is the oxide thickness and $V_{FB}^{exp}(T)$ represents the experimental temperature-dependent flat band voltage, extrapolated from Fig. 2b, that includes the interface states and the fixed charges contributions.

As shown in Fig. 5 the Fowler-Nordheim (FN) tunneling well describes the current conduction mechanisms of the system. In fact, it has been demonstrated that the ideal FN tunneling possesses a weak temperature dependence, related to the shrinking of the 4H-SiC and $SiO_2$ band gaps with increasing temperature [22]. FN tunneling is commonly achieved in $SiO_2$/Si system using thermally grown oxide layers. On the other hand, thermally grown oxide layers onto 4H-SiC contain residual carbon atom content that may affect the insulating properties of the gate oxide. In particular, Sometani et al. [11] demonstrated that thermally grown $SiO_2$ at 1200°C onto 4H-SiC shows a Poole-Frenkel (PF) emission ruling the conduction through the gate oxide in MOS capacitors at temperature above 25°C. More recently, P. Samanta and K. C. Mandal [22] compared the experimental temperature dependence of the electron barrier height, measured in MOS capacitors having a layer of $SiO_2$ thermally grown at 1100°C onto 4H-SiC, with the theoretical value of the $d\Phi_B/dT$ for the ideal FN tunneling, and observed a significant difference. In particular, the experimental slope (-7.6meV/°C) is one order of magnitude higher than the theoretical one.

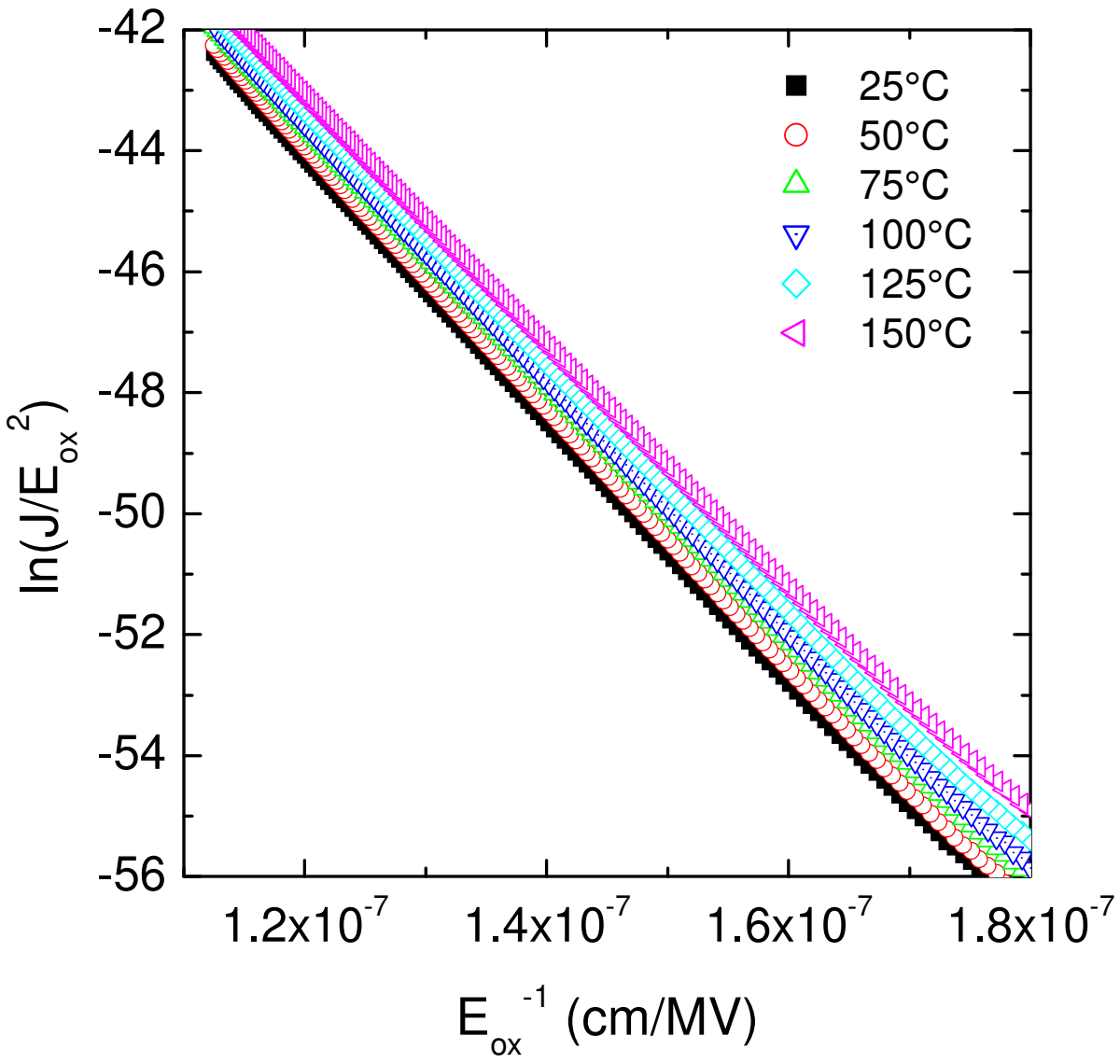

Fig. 4. Temperature dependent $\ln(J/E_{ox}^2)$-$E_{ox}^{-1}$ curves from room temperature up to 150°C

Fig. 5 shows how both our experimental results, literature data and the ideal values of the electron barrier height $\Phi_B$ vary with a temperature increasing temperature from -25°C up to 250°C. The ideal FN shows a barrier height $\Phi_B$ = 2.7 eV that decreases with a slope of the $d\Phi_B/dT$ = – 0.7meV/°C [22] with increasing temperature, essentially due to the temperature related variation of both the $SiO_2$ and 4H-SiC band offset and the Fermi level. On the other hand, an experimental barrier height of $\Phi_B$ = 2.8 eV was extracted at room temperature, i.e., slightly larger of the ideal value expected for $SiO_2$/4H-SiC interface. This result is consistent with the effective negative charge detected with the C-V curves in Fig. 2a. Furthermore, the indication of the good quality of the deposited oxide is obtained considering that the experimental slope of the $d\Phi_B/dT$ is – 0.98meV/°C is only slightly larger than the ideal value of – 0.7meV/°C. Indeed, this discrepancy is almost negligible considering a possible error of 5% in the determination of the barrier height.

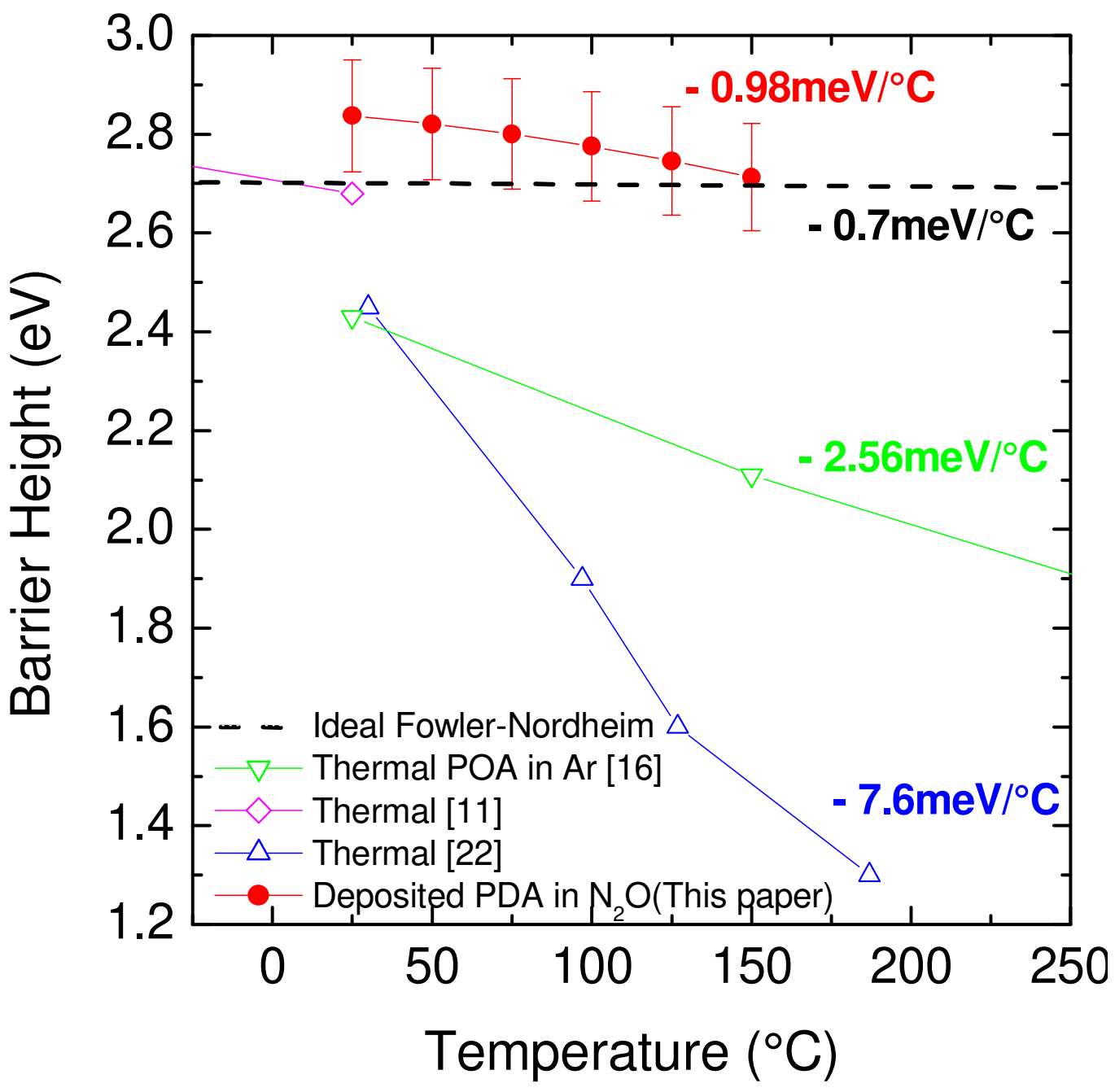

Fig. 5. Comparison between the experimental electron barrier height for the deposited (this paper), thermally grown oxides and the theoretical behavior.

In order to exclude the occurrence of a PF emission on our samples, the temperature and electric field dependence of PF emission has been considered according to the following PF formalism [23,24]:

$$ln(J/E_{ox}) = m(T)E_{ox}^{0.5} + b(T) \tag{3}$$

with

$$m(T) = \frac{q}{kT}\sqrt{\frac{q}{\pi\varepsilon_0\varepsilon_{SiO_2}}} \tag{4}$$

and

$$b(T) = -\frac{q\phi_t}{kT} + ln\,C \tag{5}$$

and where $\phi_t$ is the barrier height for electron emission from the trap states, $\varepsilon_{SiO2}$ is the relative dielectric permittivity at high frequency of the gate insulator ($SiO_2$), T is absolute temperature, $\varepsilon_0$ is the permittivity of free space, and k is Boltzmann's constant. It is known that the PF also owns an exponential relationship with the electron emission barrier height. However, the validity of the PF emission fitting can be proven looking at the temperature dependence of the linear coefficient m(T) obtained from the lineal fit of the *PF plots* $\ln(J/E_{ox})$ as a function of $E_{ox}^{0.5}$. Moreover, in the PF formalism the linear coefficient m(T) can be expressed as a linear function of $T^{-1}$ with the slope depending on the relative insulator permittivity at high frequency $\varepsilon_{SiO2}$. If a pure PF emission occurs the linear fit of the m(T) vs $T^{-1}$ should give the ideal $\varepsilon_{SiO2}=3.9$ value. However, in the present case the application of the mathematical procedure just described gives a misleading value of $\varepsilon_{SiO2}=80$. This last result pointed out that the mathematic of the PF emission has no physical meaning to describe our experimental data. Thus, this result suggests that the correct way to describe the experimental data is the FN formalism.

Fig. 5 shows also the $d\Phi_B/dT$ behavior taken from literature data [11,22] for thermally grown oxides that decreases with a slope up to ten times larger than the ideal FN value. Clearly, the behavior of the temperature coefficient of the barrier height can depend on the oxide processing. In fact, Sometani et al. [11] demonstrated the occurrence of an ideal FN behavior at low temperatures, i.e., from -150°C up to room temperature. Above room temperature, the gate oxide conduction was dominated by PF emission for thermally grown oxides and then the FN barrier was not estimated [11]. On the other hand, early study from A. K. Agarwal et al. [16] reported a $d\Phi_B/dT$ of about –2.56 meV/°C for a thermally grown oxide subjected to a POA in Ar. As mentioned before, this behavior has been explained with the occurrence of a PF emission [22], likely due to the presence of C-related defects in the insulator induced during the $SiO_2$ thermal growth.

In order to preliminarily prove that it is potentially advantageous for use in MOSFETs at high temperature, we have employed deposited oxide on the fabrication of lateral test MOSFETs devices.

Fig. 6 shows the transcharacteristics ($I_D$-$V_G$) of a lateral MOSFET having L=40µm and W=16µm measured at room temperature and at 200°C, respectively. As can be seen, the threshold voltage shifts toward smaller gate bias values increasing temperature but the enhanced mode operation of the transistor is preserved. Furthermore, the drain current increases with the increasing temperature demonstrating the improvements at high temperature in terms of total on resistance.

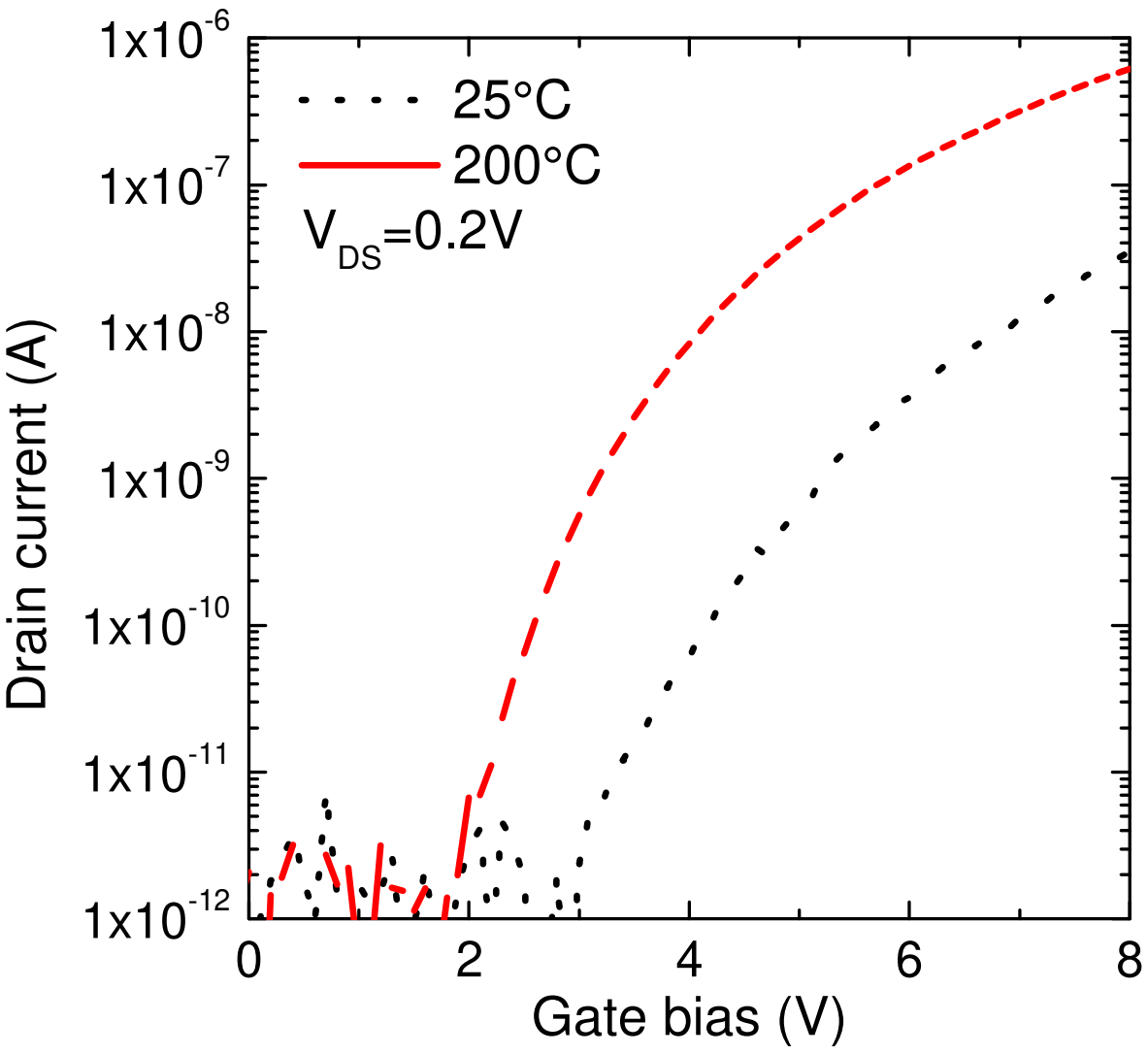

Fig. 6. Comparison between the transcharacteristics ($I_D$-$V_G$) of a lateral MOSFET having L=40µm and W=16µm measured at room temperature (dotted line) and at 200°C (dashed line).

## 4. Summary

In this paper, we have investigated the physical and temperature dependent electrical properties of $SiO_2$/4H-SiC system in MOS capacitors, with the oxide deposited by a high temperature (800°C) CVD process starting from a DCS precursor and subjected to PDA in $N_2O$. The temperature dependent electrical investigation of the MOS capacitors demonstrates the good quality of the oxide, presenting low density of fixed charges, as well as a value of $D_{it}$ comparable to our standard thermally grown oxides. The room temperature electron barrier height of 2.8 eV is close to the value expected for an ideal Fowler-Nordheim tunneling occurring between the $SiO_2$/4H-SiC conduction band. Furthermore, nearly ideal Fowler-Nordheim tunneling has been proved looking at the temperature coefficient of the electron barrier height ($d\Phi_B/dT$), which resulted noticeably improved with respect to thermally grown oxides.

**Acknowledgements**

The authors would like to thank A. La Magna and I. Deretzis for fruitful discussions. The colleagues of STMicroelectronics are acknowledged for technical assistance. In particular, F. Gentile is acknowledged for setting the thermal annealing processes, while S. Reina and A. Parisi for their valuable help during the electrical measurements.

This work was carried out in the framework of the ECSEL JU project WInSiC4AP (Wide Band Gap Innovative SiC for Advanced Power), grant agreement n. 737483.